\documentclass[journal,letterpaper]{IEEEtran}
\usepackage[cmex10]{amsmath} 
\usepackage{algorithm}
\usepackage{algorithmicx}
\usepackage{algpseudocode}
\usepackage{amssymb}
\usepackage{amsfonts}
\usepackage{arydshln} 
\usepackage{bm}
\usepackage{color}
\usepackage{graphicx}
\usepackage{multirow} 
\usepackage{soul}
\usepackage{amssymb}
\usepackage{float}
\usepackage{mathtools}
\usepackage[utf8]{inputenc}
\usepackage[english]{babel}
\usepackage{amsthm}
\usepackage[nopostdot,nogroupskip,nonumberlist]{glossaries}
\usepackage{cite}
\usepackage{lipsum}
\usepackage{clipboard}
\usepackage{booktabs}
\usepackage{float}
\usepackage{subfig}
\newclipboard{clipboard_tsp2020}
\usepackage[utf8]{inputenc} 
\usepackage[T1]{fontenc}
\usepackage{url}
\usepackage{ifthen}
\newacronym{CACC}{CACC}{cooperative adaptive cruise control}
\newacronym{dCACC}{dCACC}{degraded cooperative adaptive cruise control}
\newacronym{ACC}{ACC}{adaptive cruise control}
\newacronym{CGKF}{CGKF}{constant gain Kalman filter}
\newacronym{ZOH}{ZOH}{zero-order-hold}
\newacronym{MSE}{MSE}{mean square error}
\newacronym{MMSE}{MMSE}{minimum mean square error}
\newacronym{MIMO}{MIMO}{multiple-input-multiple-output}
\newacronym{ISAC}{ISAC}{integrated sensing and communication}
\newacronym{BS}{BS}{base station}
\newacronym{iid}{i.i.d}{independent and identically distributed}
\newacronym{DD}{DD}{data-dependent}
\newacronym{DI}{DI}{data-independent}
\newacronym{wrt}{w.r.t.}{with respect to}
\newacronym{SVD}{SVD}{singular value decomposition}
\newacronym{DFT}{DFT}{discrete Fourier transform}
\newacronym{SAA}{SAA}{sample average approximation}
\newacronym{UE}{UE}{user equipment}
\newacronym{ADC}{ADC}{analog-to-digital converter}
\newacronym{TIR}{TIR}{target impulse response}
\newacronym{RIS}{RIS}{reconfigurable intelligent surface}
\newacronym{ULA}{ULA}{uniform linear array}
\newacronym{UPA}{UPA}{uniform planar array}
\newacronym{LoS}{LoS}{line-of-sight}
\newacronym{NLoS}{NLoS}{non-line-of-sight}
\newacronym{CRB}{CRB}{Cramér-Rao bound}
\newacronym{DMA}{DMA}{dynamic metasurface antenna}
\newacronym{RCS}{RCS}{radar cross-section}
\newacronym{SINR}{SINR}{signal-to-interference-plus-noise ratio}
\newacronym{SNR}{SNR}{signal-to-noise ratio}
\newacronym{SDR}{SDR}{semidefinite relaxation}
\newacronym{NCCS}{NCCS}{no cross-correlation suppression}
\newacronym{FFBF}{FFBF}{far-field beamforming}

\usepackage[switch]{lineno}
\usepackage[bookmarks,colorlinks,bookmarksnumbered]{hyperref}
\hypersetup{colorlinks=true,pdfauthor=author}

\DeclarePairedDelimiter\abs{\lvert}{\rvert}%
\DeclarePairedDelimiter\norm{\lVert}{\rVert}%

\newcommand\myvec[1]{\bm{#1}}
\newcommand\mymat[1]{\mathbf{#1}}

\newcommand\mymatt[1]{\mathbf{\tilde{#1}}}
\newcommand\mymath[1]{\mathbf{\hat{#1}}}
\newcommand\myvect[1]{\tilde{\bm{#1}}}

\newcommand\myexpect[1]{\mathbb{E}\left\{#1\right\}}
\makeatletter
\let\oldabs\abs
\def\abs{\@ifstar{\oldabs}{\oldabs*}}
\let\oldnorm\norm
\def\norm{\@ifstar{\oldnorm}{\oldnorm*}}
\makeatother

\makeatletter
\renewcommand{\fnum@figure}{Fig. \thefigure}
\makeatother

\newtheorem{theorem}{Theorem}
\newtheorem{lemma}{Lemma}

\newtheorem{remark}{\it{Remark}}

\let\oldtheorem\theorem
\let\endoldtheorem\endtheorem

\let\oldremark\remark
\let\endoldremark\endremark

\renewenvironment{theorem}
{\oldtheorem\normalfont}
{\endoldtheorem}
\renewenvironment{remark}
{\oldremark\normalfont}
{\endoldremark}

\graphicspath{{figures/}}

\begin{document}
\title{Near-Field Integrated Sensing and Communication for Multi-Target Indication}

\author{%
	\IEEEauthorblockN{%
		Hang Ruan\,$^{1}$, 
		Homa Nikbakht\,$^{1,2}$, 
		Ruizhi Zhang\,$^{1,3}$, 
		Honglei Chen\,$^{4}$, 
		Yonina C. Eldar\,$^{1}$%
	}
	\\
	\IEEEauthorblockA{%
		$^{1}$Weizmann Institute of Science, 
		$^{2}$Princeton University, \\
		$^{3}$University of Electronic Science and Technology of China, 
		$^{4}$MathWorks, Inc.\\
		Emails: 
		\{hang.ruan, ruizhi.zhang, yonina.eldar\}@weizmann.ac.il, 
		homa@princeton.edu, 
		h.chen@mathworks.com
	}
		\thanks{
				This research was supported by the European Research Council (ERC) under the European Union’s Horizon 2020 research and innovation program (grant No. 101000967), by the Israel Science Foundation (grant No. 536/22), and by the Manya Igel Centre for Biomedical Engineering and Signal Processing.}
}



\maketitle
\thispagestyle{empty}

%
%
%
%

\begin{abstract}
	Integrated sensing and communication (ISAC) in the near‐field regime offers the potential to jointly support high‐rate downlink transmission and high‐resolution multi‐target detection by exploiting the spherical‐wave nature of electromagnetic propagation. In this paper, we propose a unified beamforming framework for a multi‐user multi‐target near‐field ISAC system. In this system, a multi‐antenna base station simultaneously serves multiple single‐antenna users and senses multiple point‐targets without prior knowledge of their radar cross sections. By optimizing the transmit covariance matrix, our design maximizes the minimum weighted transmit beampattern gain across all targets to ensure accurate sensing while strictly limiting inter‐target cross‐correlations and guaranteeing per‐user communication rate and total power constraints. We extend classical far‐field beampattern and cross‐correlation measures to the near‐field by incorporating both angle and range dependencies, enabling discrimination of targets along the same direction but at different distances. The resulting non‐convex program is efficiently relaxed to a semidefinite program via rank‐one lifting. We then develop a closed‐form reconstruction to recover optimal rank‐one beamformers. Numerical simulations demonstrate that our near‐field ISAC design can simultaneously resolve and serve users/targets along the same direction but at different distances, achieving significant gains over far‐field and single‐target benchmarks.
\end{abstract}

\section{Introduction}
\label{sec:intro}

\Gls{ISAC} has emerged as a transformative paradigm that integrates communication and sensing functionalities into a single system, leveraging shared hardware and spectrum resources \cite{liu2022integrated}. By enabling these dual functionalities, \gls{ISAC} systems can enhance the efficiency of resource utilization, reduce hardware costs, and open new possibilities for applications such as autonomous vehicles, smart cities, and industrial automation \cite{liu2022integrated}.

In recent years, near-field communication and sensing have attracted growing interest, driven by the adoption of high-frequency mmWave systems and large-aperture \gls{MIMO} \gls{BS} antennas \cite{wang2023near}. In the near-field region, where the propagation distance is on the order of the Rayleigh distance, 
the spatial channels follow a spherical rather than planar wavefront. As a result, both communication and sensing channels depend on the users’ or targets’ angles and ranges \cite{wang2023near,galappaththige2024near}. This range-dependent behavior enables the system to distinguish objects aligned in the same direction but at different distances, substantially improving spatial resolution for both functionalities \cite{wang2023near,galappaththige2024near}. Such capabilities are essential for next-generation applications that demand high accuracy and spectral efficiency.

Most existing studies on near-field \gls{ISAC} beamforming focus on serving multiple communication users while sensing a single target \cite{wang2023near,qu2024near,zhao2024modeling,li2024near}, which severely limits performance in multi-target scenarios. To address this, \cite{galappaththige2024near} considers multi-target sensing in an \gls{ISAC} system by treating echoes from all other targets as interference and optimizes the \gls{SINR} for each individual target. However, this \gls{SINR}-centric design has two fundamental drawbacks. 
First, in \gls{MIMO} sensing signal design, reducing cross-correlation among transmit waveforms is critical for accurate multi-target indication \cite{liu2020joint, meng2024multi} and for enabling adaptive MIMO radar techniques \cite{stoica2007probing}. This requirement is not imposed in \cite{galappaththige2024near}, thereby deteriorating the capacity of multi-target indication. 
Second, the method in \cite{galappaththige2024near} assumes prior knowledge of each target’s \gls{RCS} information to compute the per‐target \glspl{SINR} to evaluate multi-target sensing performance, yet \gls{RCS} values are often unavailable in practical \gls{ISAC} deployments. In contrast, evaluating sensing performance via beam‐pattern gains and cross‐correlation levels requires only known array steering functions and is entirely agnostic to \gls{RCS} values \cite{stoica2007probing,liu2020joint,meng2024multi}.

Existing far‐field MIMO beamforming designs treat each array’s response as a function of direction only, using planar wavefront models in which the steering vector varies solely with angle. As a result, classical approaches optimize both the transmit beampattern and the cross‐correlation of waveforms over angular coordinates alone \cite{stoica2007probing,liu2020joint,meng2024multi}. In the near‐field, however, each antenna element sees a spherical wavefront whose strength and phase shift depend on both the distance to and the direction of each user or target. This means that two objects aligned along the same angle but at different ranges produce non-coherent channels
. To exploit this, we generalize the traditional beampattern and cross‐correlation criteria so that they account simultaneously for angle and range. By doing so, our beamforming can not only form narrow beams in specific directions, but also focus or place nulls at particular distances, thereby enabling separation of users or targets that lie along the same direction but at different distances. This leads to greatly enhancement in both multi‐user communications and multi‐target sensing resolution in \gls{ISAC} systems.

In this work, we propose a unified framework for joint transmit beamforming in near‐field \gls{ISAC} systems that simultaneously serves multiple downlink users and resolves multiple sensing targets without relying on prior \gls{RCS} knowledge.  By optimizing the transmitted signal covariance, our formulation maximizes the worst‐case weighted beampattern gain across all targets while enforcing strict limits on mutual cross‐correlation and guaranteeing per‐user communication rate and total power constraints.  Crucially, we extend classical far‐field beam‐pattern and cross‐correlation metrics to depend on both angles and ranges, enabling the discrimination of  objects at identical directions but different distances.  To solve the resulting non-convex program, we recast it via \gls{SDR} into a convex form and derive a rank‐one recovery procedure that yields optimal beamformers in closed form. Numerical results demonstrate that our method can simultaneously distinguish and serve users and sensing targets located in the same direction but at different distances.

\textbf{Notations}: Throughout the paper, $(\cdot)^T$, $(\cdot)^*$ and $(\cdot)^H$ denote transpose, conjugate and conjugate transpose, respectively; $\|\cdot\|$ is the Euclidean norm; $\operatorname{Tr}(\cdot)$ is the trace operator; $\mathbb{E}\{\cdot\}$ is statistical expectation; $\delta[\cdot]$ is the Kronecker delta function; and $\mathbf I_n$ is the $n\times n$ identity matrix. Scalars, vectors and matrices are denoted by italic, bold lowercase and bold uppercase letters. 

\section{System and signal model}
\label{sec:model}
\subsection{System model}

\begin{figure}[htb]
	\centering	
		\includegraphics[width=0.65\linewidth]{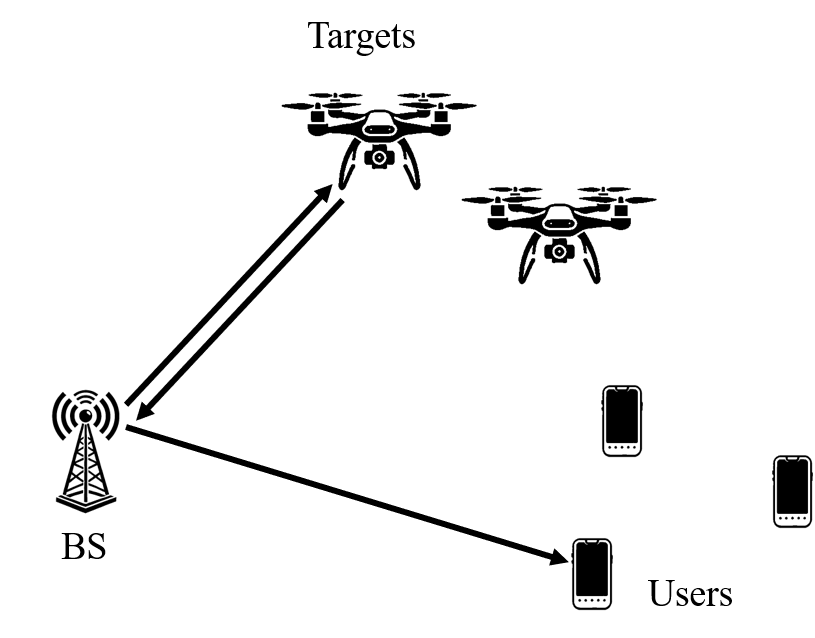}
		\label{fig:scenario_a}
	\hfill
	\caption{Illustration of the near-field \gls{ISAC} system with multiple communication users and multiple targets.}
	\label{fig:scenario_multi_target}
\end{figure}

Fig.~\ref{fig:scenario_multi_target} illustrates the considered \gls{ISAC} system, operating in the mmWave band. We assume a \gls{BS} with a planar array of \( N_t \) antennas for transmission and another planar array of \( N_r \) antennas for sensing reception. 
The position of the \( n_t \)-th transmitting antenna element (\(1 \leq n_t \leq N_t\)) is denoted as \(\myvec{p}_t(n_t) \in \mathbb{R}^3\), with the array’s radiation normal vector represented by \(\myvec{w}_t\). Analogously, the positions and radiation normal vectors for the \gls{BS}'s receiving array 
are given by \(\myvec{p}_r(n_r)\) 
and \(\myvec{w}_r\), 
respectively.

The \gls{BS} simultaneously serves downlink communication to \(K\) single-antenna users, indexed by the set \(\mathcal{K}\), and senses \(L\) targets, indexed by the set \(\mathcal{L}\). Each user/target \(a \in \mathcal{K} \cup \mathcal{L}\) has a known position \(\myvec{p}_a\)
%
%
 within the near-field region of the \gls{BS}, bounded by the Rayleigh distance \( R_{\text{RL}} = 2D^2/\lambda \) \cite{wang2023near}, where \(D\) is the antenna aperture, \(\lambda = c/f_c\) is the wavelength, \(f_c\) is the carrier frequency, and \(c\) is the speed of light.


\subsection{Transmitted \gls{ISAC} signal}
We consider a coherent time block of length $T$. In a fully-digital beamforming structure, at time $t \in \{1, \dots, T\}$, the transmitted signal of the \gls{BS} is given by \cite{hua2023optimal}
\begin{equation}
	\myvec{x}[t] = \sum\nolimits_{k \in \mathcal{K}} \myvec{f}_k c_k[t] + \myvec{s}[t],
\end{equation}
where the $n_t$-th element of $\myvec{x}[t]\in \mathbb{C}^{N_t}$ is the symbol transmitted by the $n_t$-th element ($1 \leq n_t \leq N_t$). The vector $\myvec{f}_k \in \mathbb{C}^{N_t}$ is the fully digital beamformer for communicating the symbol $c_k[t]$ to the user $k$, and $\myvec{s}[t]\in \mathbb{C}^{N_t}$ is the signal dedicated for target sensing, whose covariance is $\mymat{R}_s = \myexpect{\myvec{s}[t]\myvec{s}^H[t]}$. The communication symbols have unit power and are \gls{iid} over time and users. Thus, we have $\myexpect{c_k[t]c_l^*[t]} = \delta[k-l]$. 
The covariance of the transmitted signal $\myvec{x}[t]$ is given by
\begin{equation} \label{eqn:signal_cov}
	\mymat{R}_x = \myexpect{\myvec{x}[t]\myvec{x}^H[t]} = \sum\nolimits_{k\in\mathcal{K}} \myvec{f}_k \myvec{f}_k^H +\mymat{R}_s.
\end{equation}

\subsection{Communication model} 
\label{subsec:comm_model}
%


We employ the near-field spherical wave channel model for the channels between the \gls{BS} and users/targets. For a user \(k\in \mathcal{K}\), the channel \(\myvec{h}_{t,k}\in\mathbb{C}^{N_t}\) is modeled as \cite{zhang2022beam}:
\begin{equation} \label{eqn:comm_path_LoS}
	\left[\myvec{h}_{t,k}\right]_{n_t} = 
	\sqrt{F(\myvec{p}_k-\myvec{p}_t(n_t),\myvec{w}_t)}
	\beta(\myvec{p}_k-\myvec{p}_t(n_t)),
\end{equation}
where $F(\myvec{p}_c(k)-\myvec{p}_t(n_t),\myvec{w}_t)$ is the radiation profile of each element, with the function $F(\myvec{p},\myvec{w})$ modeled as \cite{ellingson2021path}
\begin{equation}
	F(\myvec{p},\!\myvec{w}) \! = \!
	\begin{cases}
		2(b\!+\!1) \cos^b(\psi(\myvec{p},\!\myvec{w}) ), & \psi(\myvec{p},\!\myvec{w}) \!\in\! [0, \pi/2], \\
		0, & \text{otherwise},
	\end{cases}
\end{equation}
where $\psi(\myvec{p},\myvec{w})$ denotes the angle between vectors $\myvec{p}$ and $\myvec{w}$. The parameter \( b \) determines the Boresight gain, which is set as \( b = 2 \) for the dipole case. The term $\beta(\myvec{p}_k-\myvec{p}_t(i,l))$ denotes the attenuation and phase change caused by the path, with the function $\beta(\myvec{p})$ modeled as
\begin{equation}
	\beta(\myvec{p}) = {\lambda}/\left({4\pi\norm{\myvec{p}}}\right)
	\cdot e^{-j\frac{2\pi}{\lambda} \norm{\myvec{p}}}.
\end{equation}

Here, we compare the near‐field model \eqref{eqn:comm_path_LoS} to its far‐field approximation.  In the far‐field case, the radiation profile and the path‐loss term are both taken as constants across the array, i.e.,
\(
\sqrt{F(\myvec{p}_k-\myvec{p}_t(n_t),\myvec{w}_t)}\;\approx\;\sqrt{F(\myvec{p}_k-\myvec{p}_{t,0},\myvec{w}_t)}\), 
\(
{\lambda}/({4\pi\norm{{\myvec{p}_k-\myvec{p}_t(n_t)}}})\;\approx\;{\lambda}/({4\pi\norm{{\myvec{p}_k-\myvec{p}_{t,0}}}}),
\)
and the phase term uses the distance approximation as follows:
\begin{equation}
	\label{eqn:far_field_distance}
	\begin{aligned}
		\bigl\|\myvec{p}_k-\myvec{p}_t(n_t)\bigr\|
		&\approx &&
		\bigl\|\myvec{p}_k-\myvec{p}_{t,0}\bigr\|
		-
		\bigl\|\myvec{p}_t(n_t)-\myvec{p}_{t,0}\bigr\|
		\\
		& &&\times
		\cos(\psi(\myvec{p}_t(n_t)-\myvec{p}_{t,0},\,\myvec{p}_k-\myvec{p}_{t,0}))\\
		& \triangleq && \Delta_{n_t,k}.		
	\end{aligned}
\end{equation}
Hence the far‐field channel becomes
\begin{equation}
	\label{eqn:far_field_channel}
	\begin{aligned}
		\bigl[\myvec{h}_{t,k}^{\text{far}}\bigr]_{n_t}
	\!=\!\sqrt{F(\myvec{p}_k\!-\!\myvec{p}_{t,0},\myvec{w}_t)}\,
			\beta(\myvec{p}_k-\myvec{p}_{t,0})
		\;
		e^{-j\frac{2\pi}{\lambda}\,\Delta_{n_t,k}}.
	\end{aligned}
\end{equation}
In the far-field model \eqref{eqn:far_field_channel}, the amplitude is a constant across all elements and the inter‐element phase‐differences depend only on the common directions in $\Delta_{n_t,k}$ and not on the distances. Consequently, far-field beamforming cannot resolve two users sharing the same angle but at different distances. In contrast, the near‐field model \eqref{eqn:comm_path_LoS} depends on the distance, indicating the capacity of distinguishing such users.

The received signal at user $k$ at time $t$ is
\begin{equation}
	y_k[t] = \myvec{h}_{t,k}^T \,\myvec{x}[t] + n_k[t],
\end{equation}
where $n_k[t] \sim \mathcal{CN}(0,\sigma_k^2)$ is the additive noise.
Thus, the \gls{SINR} at user \( k \) is given by:
\begin{equation} \label{eqn:comm_SINR}
	\eta_k = \frac{\abs{\myvec{h}_{t,k}^T \myvec{f}_k}^2}{\sum_{k' \in \mathcal{K}, k' \neq k} \abs{\myvec{h}_{t,k}^T \myvec{f}_{k'}}^2 + \text{Tr}(\myvec{h}_{t,k}^T \mymat{R}_s \myvec{h}_{t,k}) + \sigma_k^2}.
\end{equation}
The achievable communication rate for user \( k \) (in bits per second per Hz) is used as the communication performance metric, given by the Shannon capacity formula \cite{zhang2022beam}:
\begin{equation} \label{eqn:shannon_capacity}
	R_k = \log_2 (1 + \eta_k).
\end{equation}

\subsection{Sensing model}
The sensing channel involves round-trip propagation to each target \(l\in\mathcal{L}\), which can be separated into the forward channel (from the \gls{BS} to the target) and the backward channel (from the target to the \gls{BS}). Similarly to the near-field communication model \eqref{eqn:comm_path_LoS}, the forward channel \(\myvec{h}_{t,l}\in\mathbb{C}^{N_t}\) is modeled as
\begin{equation} \label{eqn:sensing_path_LoS}
	\left[\myvec{h}_{t,l}\right]_{n_t} = 
	\sqrt{F(\myvec{p}_l-\myvec{p}_t(n_t),\myvec{w}_t)}
	\beta(\myvec{p}_l-\myvec{p}_t(n_t)),
\end{equation}
and the backward channel \(\myvec{h}_{l,r}\in\mathbb{C}^{N_r}\) is modeled as
\begin{equation} \label{eqn:sensing_path2_LoS}
	\bigl[\myvec{h}_{l,r}\bigr]_{n_r}
	= \sqrt{F(\myvec{p}_r(n_r) - \myvec{p}_l, \myvec{w}_r)}
	\beta(\myvec{p}_r(n_r) - \myvec{p}_l).
\end{equation}
%
%
%
Thus, the overall round-trip sensing channel of target $l$, denoted as $\mymat{H}_l \in \mathbb{C}^{N_r \times N_t}$ 
is given by
\begin{equation} \label{eqn:sense_channel_non_RIS}
	\mymat{H}_l = \gamma_l \myvec{h}_{l,r} \myvec{h}_{t,l}^T,
\end{equation}
where $\gamma_l$ is the \gls{RCS} of target $l$. 
Summing up the echoes of all targets, the received sensing signal at the \gls{BS} is given by
\begin{equation} \label{eqn:sensing_signal}
	\myvec{y}_r[t] =
	\sum\nolimits_{l\in \mathcal{L}}\mymat{H}_l \myvec{x}[t] + \myvec{n}_r[t],
\end{equation}
where $\myvec{n}_r[t] \sim \mathcal{CN}(\mathbf{0}, \sigma_r^2 \mymat{I}_{N_r})$ is the sensing receiver's noise.

Similarly to the communication case, the far‐field sensing model approximates all array elements as equidistant from a given target and ignores range dependence, making it incapable of resolving two targets sharing the same direction but located at different distances. In contrast, the near‐field model in \eqref{eqn:sense_channel_non_RIS} retains the full distance dependence of each transmit‐receive path, thereby providing the angular‐and‐range resolution necessary to discriminate targets with identical directions but distinct distances.

For multi‐target MIMO sensing, the transmit covariance $\mymat{R}_x$ must be shaped to achieve two complementary objectives \cite{liu2020joint,meng2024multi,stoica2007probing}:
1)~\textbf{Beam‐pattern gain}: maximize the power delivered to each target $l$, quantified by
	\(
	\myvec{h}_{t,l}^T \,\mymat{R}_x\, \myvec{h}_{t,l}^*;
\)
2)~\textbf{Cross‐correlation suppression}: minimize mutual interference between any two targets $l\neq l'$, quantified by $\bigl|\myvec{h}_{t,l}^T \,\mymat{R}_x\, \myvec{h}_{t,l'}^*\bigr|$.


{\textbf{Remark}: In multi-target sensing, existing far‐field \gls{MIMO} beamforming methods optimize beam‐pattern and cross‐correlation metrics exclusively over angular directions, treating each array’s response as a planar wavefront that depends only on the angle \cite{stoica2007probing,liu2020joint,meng2024multi}. More recent near‐field study \cite{galappaththige2024near} extends beamforming to near-field spherical‐wave channels but focuses on \gls{SINR}‐based designs that do not jointly shape angle‐and‐range beampatterns and suppress cross‐correlation among multiple targets. As shown in Section \ref{sec:result}, neglecting cross‐correlation suppression makes classical \gls{MIMO} techniques, e.g., Capon spectrum, incapable of effectively indicating multiple targets. In contrast, our work is the first to generalize both beampattern and cross‐correlation criteria to the full near‐field model, which accounts simultaneously for angle and distance; Thus it can resolve targets along the same direction but at different distances.}

\section{Beamforming optimization}
\label{subsec:bf_opt}

We formulate the joint beamforming design for multi‐user communication and multi‐target sensing as
\begin{subequations} \label{eqn:bf_opt_non_RIS_multi_target}
	\begin{align}
		\max_{\{\!\myvec{f}_k\!\}, 
			 \mathbf{R}_s\succeq 0, \mymat{R}_x, \mu} \quad & \mu \label{eqn:opt_obj_multi_target}\\
		\text{s.t.} \quad & w_l (\myvec{h}_{t,l}^T \mymat{R}_x \myvec{h}_{t,l}^*) \geq \mu, \forall l \in \mathcal{L}, \label{eqn:opt_const_sensing_bp_multi_target}\\
		& w_{l,l'} \!\abs{\myvec{h}_{t,l}^T \mymat{R}_x \myvec{h}_{t,l'}^*\!} \!\leq\! \varepsilon \mu,  \forall l, \! l'\!\in\!\mathcal{L}, l\!\neq\!l', \label{eqn:opt_const_cross_multi_target}\\
		& R_k \geq R_{\min,k}, \quad \forall k \in \mathcal{K}, \label{eqn:opt_const_comm_multi_target}\\
		& \operatorname{Tr}(\mathbf{R}_x) \leq P_{\max}, \label{eqn:opt_const_bs_power_multi_target} \\
		& \mymat{R}_x = \sum\nolimits_{k\in\mathcal{K}} \myvec{f}_k \myvec{f}_k^H +\mymat{R}_s. \label{eqn:opt_const_cov}
	\end{align}
\end{subequations}

The optimization in \eqref{eqn:bf_opt_non_RIS_multi_target} maximizes an auxiliary variable $\mu$, which represents the worst‐case weighted sensing beampattern gain across all targets (objective \eqref{eqn:opt_obj_multi_target}), subject to the following: each target’s weighted beampattern gain must be at least $\mu$ (constraint \eqref{eqn:opt_const_sensing_bp_multi_target}); the magnitude of the cross‐correlation between any two targets, scaled by weight $w_{l,l'}$, must not exceed $\varepsilon\,\mu$ to suppress inter‐target interference (constraint \eqref{eqn:opt_const_cross_multi_target}); each user’s downlink rate must meet its minimum requirement $R_{\min,k}$ (constraint \eqref{eqn:opt_const_comm_multi_target}); the total transmit power must not exceed the budget $P_{\max}$ (constraint \eqref{eqn:opt_const_bs_power_multi_target}); and the overall transmit covariance $\mathbf{R}_x$ must equal the sum of the communication beamformer covariances and the sensing covariance as defined in \eqref{eqn:signal_cov} (constraint \eqref{eqn:opt_const_cov}).

The optimization problem \eqref{eqn:bf_opt_non_RIS_multi_target} is non-convex because of the
quadratic term $\myvec{f}_k \myvec{f}_k^H$ in the equality constraint \eqref{eqn:opt_const_cov}. To solve this optimization, we recast it into a convex optimization using the \gls{SDR} strategy following the two steps below. 

First, using the Shannon capacity formula \eqref{eqn:shannon_capacity} and the definition of \gls{SINR} \eqref{eqn:comm_SINR}, \eqref{eqn:opt_const_comm_multi_target} is equivalent to \cite{wang2023near}
\begin{equation}\label{eqn:constraint-in-trace-form}
	\operatorname{Tr}\Bigl(
	\myvec{h}_{t,k}^{\ast}\,\myvec{h}_{t,k}^{T}
	\bigl(\,
	\xi_{k}\,\myvec{f}_{k}\,\myvec{f}_{k}^{H}
	-
	\mathbf{R}_{x}
	\bigr)
	\Bigr)
	\ge
	\sigma_{k}^{2},
\end{equation}
where $\xi_{k} =
{2^{\,R_{\min,k}}}/({2^{R_{\min,k}} - 1})$.
Second, we replace the optimized variable $\myvec{f}_k$ with the matrix $\mymat{F}_k =  \myvec{f}_k \myvec{f}_k^H$ and leave out the constraint that $\mymat{F}_k$ is a rank-one matrix. Thus, the optimization \eqref{eqn:bf_opt_non_RIS_multi_target} is relaxed to
\begin{subequations} \label{eqn:bf_opt_non_RIS_multi_target2}
	\begin{align}
		\max_{\{ \!\mymat{F}_k\! \}, 
			\mathbf{R}_s\succeq 0, \mymat{R}_x, \mu} \quad & \mu \label{eqn:opt_obj_multi_target2}\\
		\text{s.t.} \quad & w_l (\myvec{h}_{t,l}^T \mymat{R}_x \myvec{h}_{t,l}^*) \geq \mu, \  \forall l \in \mathcal{L}, \label{eqn:opt_const_sensing_bp_multi_target2}\\
		& w_{l,l'} \!\abs{\myvec{h}_{t,l}^T \mymat{R}_x \myvec{h}_{t,l'}^*\!} \!\leq\! \varepsilon \mu,  \forall l, \! l'\!\in\!\mathcal{L}, l\!\neq\!l'\!, \label{eqn:opt_const_cross_multi_target2}\\
		& \operatorname{Tr}\left(
		\myvec{h}_{t,k}^{\ast}\myvec{h}_{t,k}^{T}
		(\xi_{k} \mymat{F}_k
		\!-\!
		\mathbf{R}_{x})
		\right)
		\!\ge\!
		\sigma_{k}^{2}, \forall k \!\in\! \mathcal{K}, \label{eqn:opt_const_comm_multi_target2}\\
		& \operatorname{Tr}(\mathbf{R}_x) \leq P_{\max}, \label{eqn:opt_const_bs_power_multi_target2} \\
		& \mymat{R}_x = \sum\nolimits_{k\in\mathcal{K}} \mymat{F}_k +\mymat{R}_s. \label{eqn:opt_const_cov2}
	\end{align}
\end{subequations}

The relaxed optimization \eqref{eqn:bf_opt_non_RIS_multi_target2} is convex and can be solved with cvx in MATLAB. The optimum is denoted as $\{\mymath{F}_k\}_{k\in \mathcal{K}}$ and $\mymath{R}_s$. We can construct the following rank-one solution that achieves the same objective value without violating the constraints of \eqref{eqn:bf_opt_non_RIS_multi_target} \cite[Theorem 1]{liu2020joint}:
\begin{equation}
	\myvect{f}_k = \left(\myvec{h}_{t,k}^T \mymath{F}_k \myvec{h}_{t,k}^*\right)^{-1/2} \mymath{F}_k \myvec{h}_{t,k}^*.
\end{equation}
{\begin{proof}
	First, by \cite[Theorem 1]{liu2020joint}, \(\myvect{f}_k\myvect{f}_k^H \preceq \mymath{F}_k\). Hence if we set
	\[
	\mymatt{R}_s \;=\; \mymath{R}_s \;+\;\sum_{k=1}^K\Bigl(\mymath{F}_k - \myvect{f}_k\myvect{f}_k^H\Bigr),
	\]
	then $\mymatt{R}_s\succeq \mymat{0}$ and the total transmit covariance based on $\{\myvect{f}_k\}$ and $\mymatt{R}_s$ is unchanged. Thus, the constraints \eqref{eqn:opt_const_sensing_bp_multi_target}, \eqref{eqn:opt_const_cross_multi_target} and \eqref{eqn:opt_const_bs_power_multi_target} remain satisfied.
	Moreover, \cite[Theorem 1]{liu2020joint} also shows that \(\bigl|\myvec{h}_{t,k}^T\myvect{f}_k\bigr|^2 = \myvec{h}_{t,k}^T\,\mymath{F}_k\,\myvec{h}_{t,k}^*\), 
	so all the SINR constraints \eqref{eqn:constraint-in-trace-form} remain satisfied, ensuring that the rate constraints  \eqref{eqn:opt_const_comm_multi_target} are satisfied. Therefore \(\{\myvect{f}_k\}\) and $\mymatt{R}_s$ form a feasible rank-one solution achieving the same objective.
\end{proof}}

\section{Simulation results}
\label{sec:result}
{In this section, we evaluate the performance of the proposed near‐field ISAC beamforming design via numerical simulations. Specifically, we compare our method against two benchmarks: \gls{NCCS}, which omits the cross‐correlation suppression constraint, and \gls{FFBF}, which employs a far‐field beamforming model. We will demonstrate gains in both downlink communication (SINR and achievable rate) and multi‐target sensing (Capon spectrum) under mmWave configurations.}

\subsection{Parameter and method configurations}
\label{subsec:config}
The operating frequency is set to $ f_c = 30 $ GHz, which corresponds to a wavelength of $ \lambda = 10 $ mm. The transmit power at the \gls{BS} is set to $ P_{\max} = 23 $ dBm, while the noise power is configured as $ \sigma^2 = -80 $ dBm for the \gls{BS} receiver and all the communication users. The block length is $T = 1000$.

The transmitting and receiving arrays of the \gls{BS} are both \glspl{UPA} with \( N_t = 10\times 10 \) elements, where the inter-element spacing is half of the signal wavelength, \( \lambda/2 \). The corresponding Rayleigh distance of near-field region is $R_{\text{RL}} = 1$ m, 
where the antenna aperture is given by $ D = 0.071 $ m. 
The \gls{BS} transmitter and receiver are centered at $ \myvec{p}_t = (0,0,0) $ m and $ \myvec{p}_t = (0,0.06,0) $ m,  whose radiation normal vectors are both $\myvec{w}_t = (0, 0, 1)$.

The number of communication users is $K=2$, indexed as User 1 and 2, whose positions are $(0,0.1,0.1) $ m and $(0,0.5,0.5)$ m, respectively. 
The number of sensing targets is $L=3$, indexed as Target 1, 2 and 3, whose positions are $ (0, -0.1, 0.1) $ m, $ (0, -0.25, 0.25) $ m and $(0,0,0.4)$ m, respectively. The \glspl{RCS} of targets are all $1$. Note that User 1 and User 2 locate along the same direction from the \gls{BS} transmitter but at different ranges, as do Targets 1 and 2.

In the optimization \eqref{eqn:bf_opt_non_RIS_multi_target}, we set a minimum rate requirement of $ R_{\min} = 17 $ bps/Hz for both users. In beam-pattern and cross-correlation constraints, i.e., \eqref{eqn:opt_const_sensing_bp_multi_target} and \eqref{eqn:opt_const_cross_multi_target}, we set the weights as $w_{l} = 1/\norm{\myvec{h}_{t,l}}^2$ and $w_{l,l'} = 1/(\norm{\myvec{h}_{t,l}}\cdot\norm{\myvec{h}_{t,l'}})$, respectively, to balance the sensing requirements across all targets regardless of their individual channel strengths. The cross‐correlation tolerance parameter is $\varepsilon = 0.1$.

In addition, we consider two benchmark schemes for comparison:
1) \textbf{\Gls{NCCS}}: 
	Here we solve the same problem as \eqref{eqn:bf_opt_non_RIS_multi_target} but omit constraint \eqref{eqn:opt_const_cross_multi_target}, thus isolating the impact of cross‐correlation suppression on multi‐target sensing.
2) \textbf{\Gls{FFBF}}: 
	To highlight the benefits of near‐field design, we resolve \eqref{eqn:bf_opt_non_RIS_multi_target} after replacing every channel vector $\myvec{h}_{t,a}$ ($a\in\mathcal{K}\cup\mathcal{L}$) with its far‐field approximation from \eqref{eqn:far_field_channel}. Additionally, since far‐field users or targets at the same direction become fully coherent, we relax the communication constraint by lowering $R_{\min}$ to 0.95 bps/Hz and remove  the cross-correlation constraint \eqref{eqn:opt_const_comm_multi_target} for the target pair 
	$(1,2)$, thereby ensuring problem feasibility. Note that if we set $R_{1,\min} = R_{2,\min} = R_{\min}$, the optimization is feasible only if $R_{\min}<1$ bps/Hz since the communication channels for the users are fully coherent in far-field modeling.

\subsection{Communication performance}
\label{subsec:results_comm}
\begin{figure}[htb]
	\centering
	\subfloat[User 1’s received power over the \(yz\)–plane]{
		\includegraphics[width=0.9\linewidth]{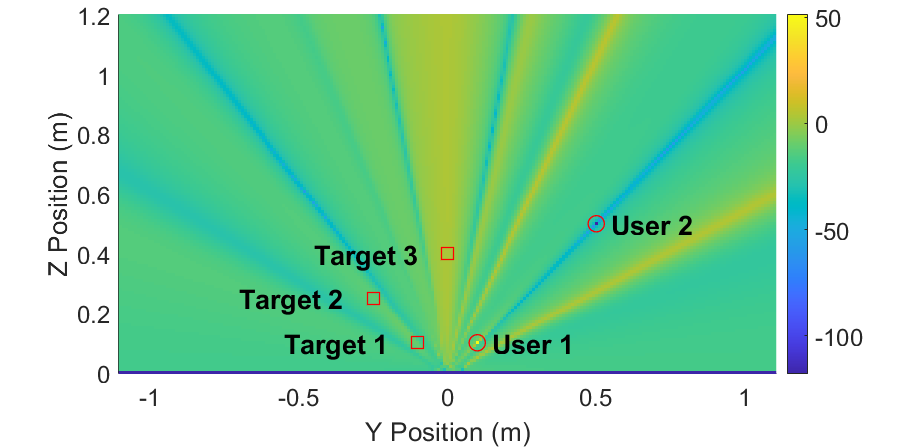}
		\label{fig:comm_SINR1_alg1}
	}
	\hfill
	\subfloat[User 2’s received power over the \(yz\)–plane]{
		\includegraphics[width=0.9\linewidth]{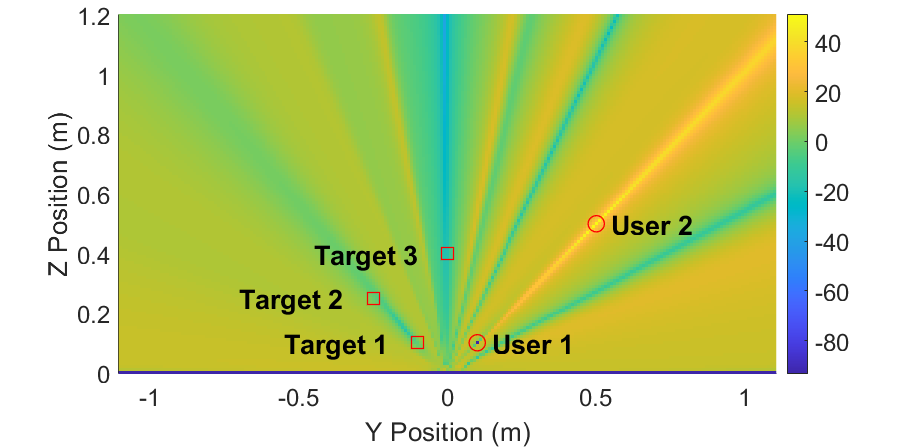}
		\label{fig:comm_SINR2_alg1}
	}
	\caption{\Gls{SINR} (in dB) of each user’s downlink signal over the $yz$-plane with proposed method.}
	\label{fig:comm_SINR_alg1}
\end{figure}

\begin{figure}[htb]
	\centering
	\subfloat[User 1’s received power over the \(yz\)–plane]{
		\includegraphics[width=0.9\linewidth]{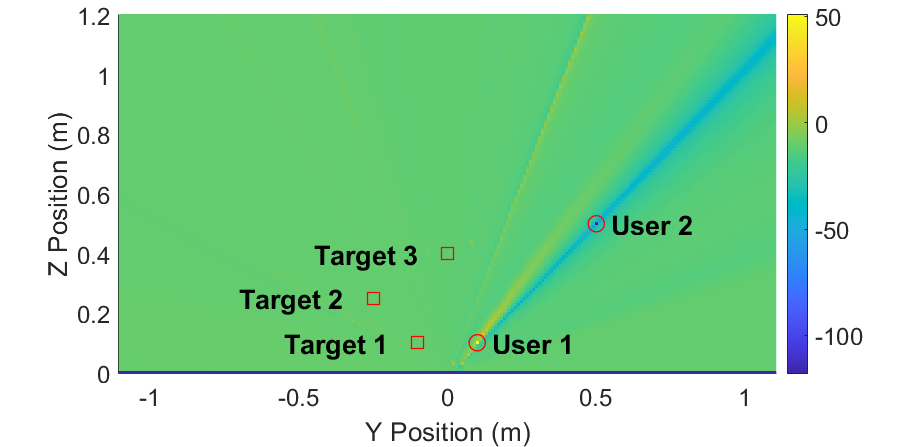}
		\label{fig:comm_SINR1_alg2}
	}
	\hfill
	\subfloat[User 2’s received power over the \(yz\)–plane]{
		\includegraphics[width=0.9\linewidth]{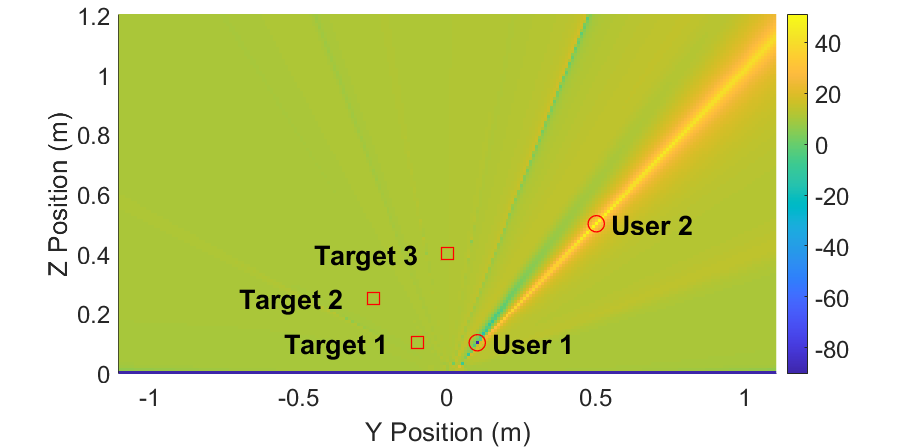}
		\label{fig:comm_SINR2_alg2}
	}
	\caption{\Gls{SINR} (in dB) of each user’s downlink signal over the $yz$-plane with \gls{NCCS}.}
	\label{fig:comm_SINR_alg2}
\end{figure}

\begin{figure}[htb]
	\centering
	\subfloat[User 1’s received power over the \(yz\)–plane]{
		\includegraphics[width=0.9\linewidth]{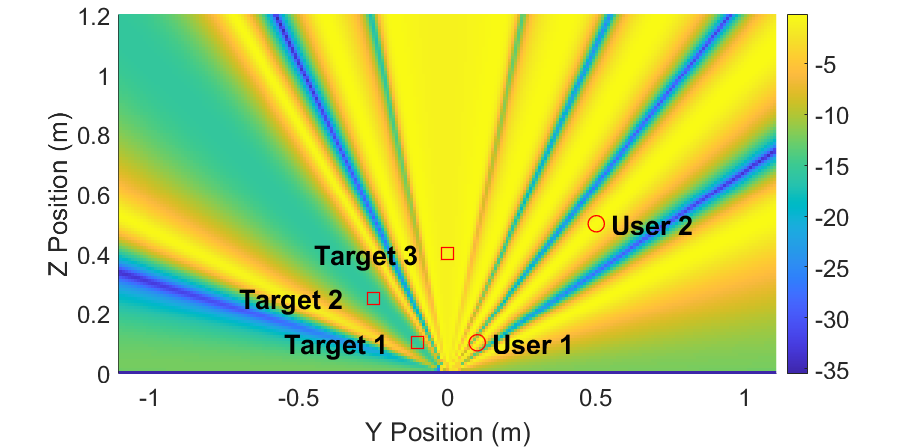}
		\label{fig:comm_SINR1_alg3}
	}
	\hfill
	\subfloat[User 2’s received power over the \(yz\)–plane]{
		\includegraphics[width=0.9\linewidth]{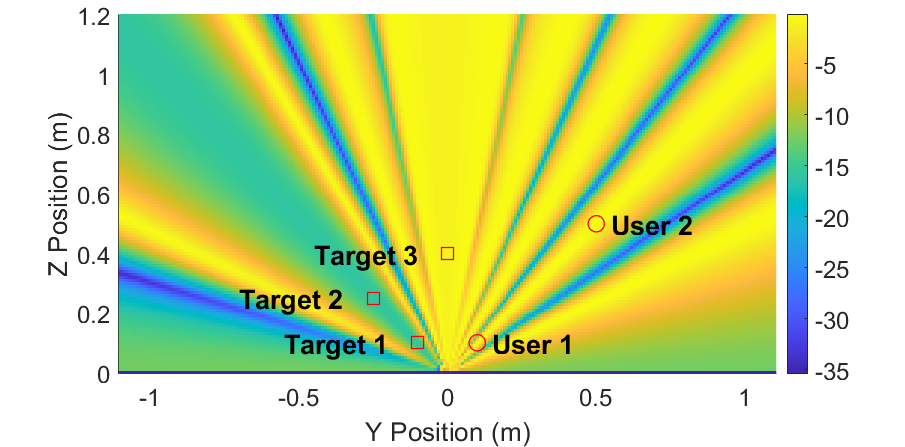}
		\label{fig:comm_SINR2_alg3}
	}
	\caption{\Gls{SINR} (in dB) of each user’s downlink signal over the $yz$-plane with \gls{FFBF}.}
	\label{fig:comm_SINR_alg3}
\end{figure}

To showcase the spatial discrimination enabled by our near-field design, Fig.~\ref{fig:comm_SINR_alg1}, \ref{fig:comm_SINR_alg2} and \ref{fig:comm_SINR_alg3} plot each user’s \gls{SINR} over the entire \(yz\)–plane under the proposed method, \gls{NCCS}, and \gls{FFBF}, respectively.  In both the proposed and \gls{NCCS} schemes (Fig.~\ref{fig:comm_SINR_alg1}, \ref{fig:comm_SINR_alg2}), each user’s \gls{SINR} exhibits a sharp peak at its location and deep nulls at the other user's location, confirming the ability to resolve users aligned in the same direction but at different distances.  By contrast, the far‐field beamforming in Fig.~\ref{fig:comm_SINR_alg3} produces overlapping \gls{SINR} patterns that cannot distinguish users along the same direction but at different distances; Moreover, the peak \gls{SINR} remains below 0 dB, which is far weaker than the above 40 dB peaks achieved by the near-field approaches. Such \gls{SINR} level renders reliable communication impossible, thus highlighting the advantage of near-field \gls{ISAC} beamforming for multi‐user separation.

\subsection{Sensing performance}
In this section, we examine the localization performance
of different beamforming methods with the Capon spectrum. Specifically, after the beamforming design, we generate the sensing signal using \eqref{eqn:sensing_signal}. Then, for any position $\myvec{p}_0$, the normalized Capon spectrum is given by \cite{xu2008target}
\begin{equation} \label{eqn:capon}
	\beta_{\text{capon}} (\myvec{p}_0) = \frac{\myvect{h}_{0,r}^H \mymat{R}_Y^{-1} \mymat{Y} \mymat{X}^H \myvect{h}_{t,0}^*}
	{T(\myvect{h}_{0,r}^H \mymat{R}_Y^{-1} \myvect{h}_{0,r}) (\myvect{h}_{l,0}^T \mymat{R}_X^{-1} \myvect{h}_{l,0}^*)}.
\end{equation}
Here $\mymat{X}=[\myvec{x}[1],\dots,\myvec{x}[T]]$ and $\mymat{Y}=[\myvec{y}[1],\dots,\myvec{y}[T]]$ collect the transmitted and received snapshots, and $\mymat{R}_X=\frac1T\mymat{X}\mymat{X}^H$, $\mymat{R}_Y=\frac1T\mymat{Y}\mymat{Y}^H$ are their sample covariance matrices. The normalized steering vectors $\myvect{h}_{t,0}=\myvec{h}_{t,0}/\|\myvec{h}_{t,0}\|$ and $\myvect{h}_{0,r}=\myvec{h}_{0,r}/\|\myvec{h}_{0,r}\|$ are formed by evaluating \eqref{eqn:sensing_path_LoS} and \eqref{eqn:sensing_path2_LoS} at $\myvec{p}_0$ and then normalizing. Compared to the conventional Capon spectrum, \eqref{eqn:capon} employs normalized vectors to remove range‐dependent amplitude scaling.
\begin{figure}[htb]
	\centering
	\subfloat[Normalized Capon spectrum (in dB) with proposed method]{
		\includegraphics[width=0.8\linewidth]{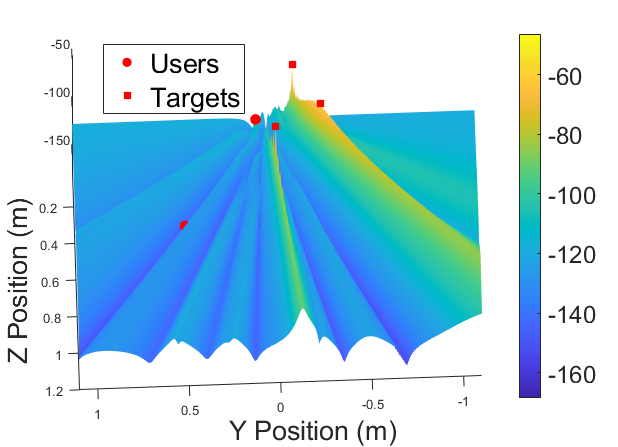}
		\label{fig:capon_alg1_sub}
	}
	\hfill
	\subfloat[Normalized Capon spectrum (in dB) with \gls{NCCS}]{
		\includegraphics[width=0.8\linewidth]{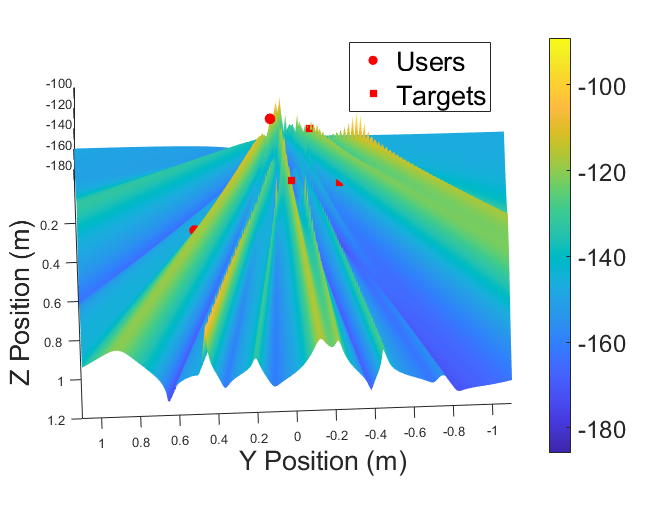}
		\label{fig:capon_alg2_sub}
	}
	\hfill
	\subfloat[Normalized Capon spectrum (in dB) with \gls{FFBF}]{
		\includegraphics[width=0.8\linewidth]{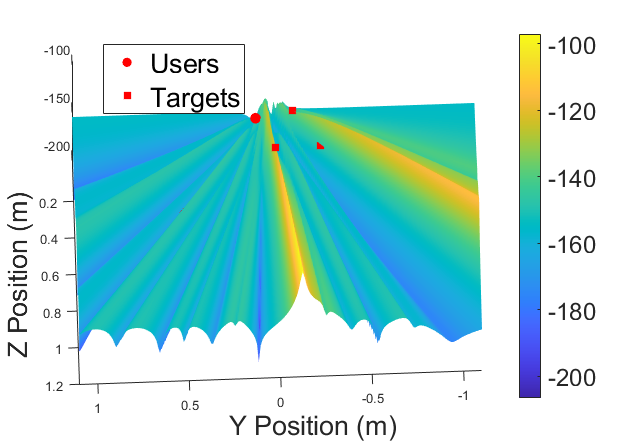}
		\label{fig:capon_alg3_sub}
	}
	\caption{Normalized Capon spectrum (in dB) of sensing over the $yz$-plane with (a) proposed method, (b) \gls{NCCS}, (c) \gls{FFBF}.}
	\label{fig:capon_algs}
\end{figure}

%
%

Fig.~\ref{fig:capon_algs} plots the normalized Capon spectrum over the entire \(yz\)–plane for the proposed method, \gls{NCCS}, and \gls{FFBF}. In Fig.~\ref{fig:capon_alg1_sub}, three distinct peaks arise exactly at the true positions of Targets 1–3, confirming that our near-field beamformer simultaneously captures both angular and range information and successfully separates targets along the same direction (Targets 1 and 2). By contrast, Fig.~\ref{fig:capon_alg2_sub} yields a cluttered spectrum with spurious peaks and no clear target signatures, illustrating that neglecting cross-correlation suppression renders classical Capon processing ineffective in multi‐target scenarios. Finally, Fig.~\ref{fig:capon_alg3_sub} cannot distinguish target distances—it shows only angular lobes—and it merges Targets 1 and 2 into a single peak. This highlights the far‐field model’s inability to resolve targets along the same direction in range or combined angle–range domains.
Overall, these results underscore the clear advantage of our joint beampattern and cross‐correlation optimization in near‐field \gls{ISAC} for precise multi‐target localization.

\section{Conclusion}
In this paper, we developed a beamforming strategy for near‐field \gls{ISAC} systems that concurrently serves multiple downlink users and senses multiple targets without requiring prior radar cross‐section information. By formulating a max–min beampattern gain optimization with explicit cross‐correlation suppression, per‐user rate guarantees, and total power limits, we extended classical far‐field metrics to jointly account for angle and range. Such an extension enabled the separation of objects along the same direction but at different distances. The original non‐convex design was relaxed into a semidefinite program, and we proposed a simple rank‐one reconstruction yielding closed‐form beamformers that attain the global optimum. Numerical results 
demonstrated that our approach substantially outperforms both no cross-correlation suppression and far‐field benchmarks in terms of communication rate and multi‐target localization accuracy. Our framework thus paves the way for practical near‐field ISAC deployments in next‐generation wireless networks.

\bibliographystyle{IEEEtran}
\bibliography{2D}








\end{document}